\documentclass{amsart}
\usepackage{amsmath,supertabular}
\usepackage{amssymb,mathrsfs}
\usepackage[hypertex,linkcolor=red]{hyperref}
\def\set#1{{\sf #1}}
\def\sH{\set{H}}\def\sK{\set{K}}
\def\sU{\set{U}}
\def\mU{\mathcal{U}}
\def\sV{\set{V}}
\def\mV{\mathcal V}
\def\sA{\set A}
\def\oA{\mathcal A}
\def\sB{\mathcal B}

\def\MO{{\operatorname M}}
\def\d{\operatorname{d}}\def\<{\langle}\def\>{\rangle}
\def\Tr{\operatorname{Tr}}\def\:{\hbox{\bf :}}
\def\openone{1\!\!1}
\def\Cmplx{\mathbb C}
\def\L#1{\set{B}(#1)}\def\St#1{\set{S(\set{#1})}}
\def\inp{\set{\,in}}\def\out{\set{\,out}}
\def\map#1{{\mathscr{#1}}}

\def\set#1{{\sf #1}}
\def\grp#1{{\mathbf #1}}
\def\transp#1{{#1}^\tau }

\def\rnk{\operatorname{rank}}
\def\Rng{\set{Rng}}
\def\Supp{\set{Supp}}\def\Span{\set{Span}}
\def\dim{\operatorname{dim}}
\def\CP #1{\set{CP(#1)}}

\def\L#1{\set {Lin}(#1)}
\def\qed{$\,\blacksquare$\par}
\def\eg{e. g. }
\def\n#1{|\!|#1|\!|}
\def\dag{\dagger}
 \def\kk{\rangle\!\rangle}\def\bb{\langle\!\langle}
\renewcommand{\geq}{\geqslant}\renewcommand{\leq}{\leqslant}
\def\ISO{{\mathbf I}} 
\newtheorem{Def}{Definition}
\newtheorem{lemma}{Lemma}
\newtheorem{Prop}{Proposition}
\newtheorem{Cor}{Corollary}
\newtheorem{Theo}{Theorem}
\def\Proof{\medskip\par\noindent{\bf Proof. }}

\begin{document}
\title[Realization schemes for quantum instruments in finite dimensions]{Realization schemes for quantum instruments in finite dimensions}

\author{Giulio Chiribella} 

\address{Giulio Chiribella, {\em QUIT} Group, Dipartimento di Fisica
  ``A. Volta'', via Bassi 6, I-27100 Pavia, Italy,
  \url{http://www.qubit.it}}\email{chiribella@unipv.it}

\author{Giacomo Mauro D'Ariano} 

\address{Giacomo Mauro D'Ariano, {\em QUIT} Group, Dipartimento di
  Fisica ``A. Volta'', via Bassi 6, I-27100 Pavia, Italy,
  \url{http://www.qubit.it}\\ Istituto Nazionale di Fisica Nucleare, Sezione di Pavia,\\
  Center for Photonic Communication and Computing, Department of
  Electrical and Computer Engineering, Northwestern University,
  Evanston, IL 60208} \email{dariano@unipv.it}

\author{Paolo Perinotti}
\address{Paolo Perinotti, {\em QUIT} Group, Dipartimento di Fisica
  ``A. Volta'', via Bassi 6, I-27100 Pavia, Italy,
  \url{http://www.qubit.it}\\ Istituto Nazionale di Fisica Nucleare,
  Sezione di Pavia}\email{perinotti@unipv.it} \date{\today}
\begin{abstract} We present a general dilation scheme for quantum instruments with continuous
  outcome space in finite dimensions, in terms of an indirect POVM measurement performed on a finite
  dimensional ancilla. The general result is then applied to a large class of instruments generated
  by operator frames, which contains group-covariant instruments as a particular case, and allows
  one to construct dilation schemes based on a measurement on the ancilla followed by a conditional
  feed-forward operation on the output.  In the case of tight operator frames our construction
  generalizes quantum teleportation and telecloning, producing a whole family of generalized
  teleportation schemes in which the instrument is realized via a joint POVM at the sender combined
  with a conditional feed-forward operation at the receiver.
\end{abstract}
\maketitle

\section{Introduction}
The theory of quantum measurements for discrete spectrum has been formulated by von Neumann in the
pioneering work \cite{vonNeumannbook}. For continuous-outcome quantum measurements, however, a
satisfactory theory has been laking for another thirty years, when the problem was finally settled
by Ozawa \cite{Ozawacontextens}. The main difficulties with the continuous-outcome measurements were
\emph{i)} the issue of repeatability, and \emph{ii)} the compatibility between the statistics  of
the measurement and the dynamical evolution of the observed system and the measuring apparatus.  In
their pioneering work \cite{Davies-Lewis-repeat}, Davies and Lewis introduced an operational
framework for the statistical description based on the mathematical concept of "instrument"---i.e.
of transformation-valued measure.  In this framework they formulated a weak repeatability
hypothesis, and conjectured that instruments for continuous-outcome measurements can never be
repeatable, even in such weak sense.  Davies and Lewis, however,  overlooked the requirement of complete
positivity of the state change in measurements, which was instead estabilished by Kraus
\cite{Kraus-annal} in the particular case of yes-no measurements.  Thirteen years later Ozawa
\cite{Ozawacontextens} showed that the state change due to an arbitrary measuring process are
described by completely positive (CP) instruments, and, viceversa, that any CP instrument can be
dilated to an indirect measurement process, with the measured system unitarily interacting with an
ancilla which then undergoes the measurement of a von Neumann observable with the same outcome space
of the instrument.  In the same paper, Ozawa finally proved the Davies and Lewis conjecture for CP
instruments, showing that they cannot be weakly repeatable unless their outcome space is discrete.

A von Neumann observable with continuous outcome space is a projection valued measure (PVM), such as
the spectral measure of a selfadjoint operator with continuous spectrum. Such an continuous-outcome
observable can exist only for infinite dimensional systems. It follows that the Ozawa dilations of
quantum instruments with continuous outcome space, even for finite dimensional systems, require an
infinite dimensional ancilla. A general positive operator valued measure (POVM), on the contrary,
can have a continuous outcome space even for finite dimensions, \eg for the measurement of the spin
direction \cite{spin}. Recently, in Refs.  \cite{continuous_povms,BarycentricPOVMs} it has been
shown that in finite dimensions every continuous-outcome POVM can be achieved as the randomization
of finite-outcome POVMs with no more than $d^2$ outcomes, $d$ being the dimension of the system's
Hilbert space.  Exploiting Naimark dilation \cite{NaimarkMA40} of the finite-outcome POVMs involved
in the randomization, this implies that for finite dimensions any continuous-outcome POVM can be
realized as a randomized observable with dimensions no greater than $d^2$.  Therefore, realization
of an instrument via indirect measurement of a POVM on a finite-dimensional ancilla allows one
to achieve the instrument as the indirect measurement of a randomized observable in finite
dimension.  

The existence of the dilation for the continuous-outcome instrument to
a POVM on a finite-dimensional ancilla has not been considered yet in
the literature, and it is not a priori obvious, since the usual
dilation procedure exploits the orthogonality of the PVM. Such an
indirect-POVM realization of the instrument is the main result of the
present paper, where we construct a general realization scheme for a
quantum instrument with continuous outcome space in finite dimension,
in terms of an indirect POVM measurement performed on an ancilla
interacting with the system. In addition, in this paper we define the
notion of instruments generated by operator frames and specialize our
dilation theorem to this case, showing that any such instrument allows
a realization in terms of an ancilla measurement followed by a
conditional feed-forward operation.  For tight operator frames, the
feed-forward scheme becomes a generalized teleportation scheme, namely
a scheme where a sender performs a joint POVM measurement on the input
system and locally on another system of an entangled pair, and
communicates the measurement outcome to the receiver, who then
performs a suitable conditional quantum channel on the other system of
the entangled pair.  The notion of instruments and channels generated
by frames and the related feed-forward realization schemes provide a
general framework encompassing a great deal of existing experimental
schemes \cite{feedforampli,feedforclon,feedforest,feedforsqeez}, and
theoretical proposals, such as tele-cloning \cite{Murao:1999p3163} and
tele-UNOT \cite{BuzEk:1999p196}.

The paper is organized as follows: in Sect. \ref{sec:gen} we recall
the preliminary notions used in the paper, also giving a new compact
rule for expressing the minimal Stinespring dilation of a given CP map
as a function of its Choi-Jamio\l kowski operator.  The general
dilation theorem for quantum instruments in finite dimensions is then
presented in Sect. \ref{sec:gendil}, where we construct an indirect
measuring process based on a POVM on a finite dimensional ancilla.  In
Sect. 4 we introduce a class of instruments generated by operator
frames, which contains group-covariant instruments as a special case.
The case of group-covariant instruments is then analyzed in detail in
Sect.  \ref{sec:covinst}. In Sect. \ref{tele} we specialize to the
case of instruments generated by tight operator frames, showing that
any such instrument can be realized via generalized teleportation
scheme, with a joint measurement at the sender and a conditional
operation at the receiver.  This construction generalizes quantum
teleportation and provides the general framework for quantum tasks
such as tele-cloning \cite{Murao:1999p3163} and tele-UNOT
\cite{BuzEk:1999p196}, as shown in Sect.  \ref{sec:covchan}.

\section{General Notions on Quantum Instruments and Channels}\label{sec:gen}

In the following we will denote by $\L{\sH}$ the vector space of
linear operators on the Hilbert space $\sH$, and by $\L{\sH, \sK}$ the
vector space of linear operators from $\sH$ to $\sK$. We will
exclusively consider finite dimensional Hilbert spaces.  Moreover, we
will denote by $\St{H}$ the convex set of density matrices on $\sH$,
and by $\CP{H,K}$ the convex cone of completely positive (CP) maps
from $\L{\sH}$ to $\L{\sK}$.

\subsection{Quantum Operations}
In Quantum Mechanics the most general evolution of a system is
described by a {\bf quantum operation}\cite{Kraus}. We will consider
generally different input and output systems in the evolution, and
denote by $\set{H_\inp}$ and $\set{H_\out}$ the corresponding Hilbert
spaces. Then, a quantum operation with input space $\set{H_\inp}$ and
output space $\set{H_\out}$ is a CP-map $\map{E}
\in\CP{H_\inp,H_\out}$ which is also trace-non-increasing. The operation $\map E$
transforms the input state $\rho_\inp\in\St{H_\inp}$ into the output
state $\rho_\out\in\St{H_\out}$ as follows
\begin{equation}
  \rho_\out=\frac{\map{E}\left( \rho_\inp\right)}{\Tr[\map{E}\left( \rho_\inp\right)]},\label{Sred}
\end{equation}
the transformation occurring with with probability
$p_{\map{E}}:=\Tr[\map{E}\left( \rho_\inp\right)]$ among a set of
possible transformations. In the deterministic case the map $\map{E}$
is trace preserving, and the quantum operation is usually called {\bf
  quantum channel}.

\subsection{Representations of operators and CP-maps}

\subsubsection{Operators and bipartite vectors} In finite dimensions it is convenient to exploit the  isomorphism
between $\L{\sH_\inp,\sH_\out}$ and $\sH_\out\otimes\sH_\inp$ induced
by the linear map
\begin{equation}\label{basis-dep}
  F\in \L{\sH_\inp, \sH_\out} \longmapsto |F \kk :=  (F \otimes \openone_\inp)  |\openone_\inp \kk  \in \sH_\out \otimes \sH_\inp~, 
\end{equation}
where $|\openone_\inp\kk \in \sH_\inp$ is the maximally entangled
vector $|\openone_\inp \kk := \sum_n |e_n\> |e_n\>$ defined by the
choice of a distinguished orthonormal basis
$\{|e_n\>\}_{n=1}^{d_\inp}$ for any copy of $\sH_\inp$.
 
Fixing an orthonormal basis $\{|c_m\>\}_{m=1}^{d_\out}$ for any copy
of $\sH_\out$, the transpose and the complex conjugate of $F$ are
uniquely defined through the relations
\begin{eqnarray}\label{tau}
  (\openone_\out \otimes F^\tau) |\openone_\out \kk &=& |F\kk\\
  F^*  &=& (F^\dag)^\tau ~, \label{star}
\end{eqnarray}
where $|\openone_\out \kk = \sum_m |c_m \> |c_m \>$, and $F^\dag \in
\L {\sH_\out, \sH_\inp}$ is the adjoint of $F$.  Definitions (\ref{basis-dep}) and (\ref{tau})  imply the elementary identities
\begin{equation}\label{kkprop}
\begin{array}{lll}
(B \otimes A) |F \kk &=& |B F A^\tau \kk  \\
\bb F | G \kk &=& \Tr[F^\dag G] := \<F, G\>_{HS} ~,     
\end{array}
\end{equation}
where  $B$ and $A$ are arbitrary operators in $\L{\sH_\out}$ and $\L{\sH_\inp}$, respectively, and $\< \cdot,\cdot\>_{HS} $ denotes the Hilbert-Schmidt scalar product in $\L{\sH_\inp, \sH_\out}$.

\subsubsection{Linear maps and bipartite operators} In finite
dimensions it is convenient to represent linear maps $\map{M}$ from
$\L{\sH_\inp}$ to $\L{\sH_\out}$ as linear operators $R_{\map M}$ on
$\sH_\out\otimes\sH_\inp$ via the so-called Choi-Jamio\l kowski (CJ)
isomorphism \cite{choi75,Jamiolkowski72}
\begin{equation}\label{R}
  R_\map{M} = (\map{M} \otimes\map{I}) \left( |\openone_\inp\kk\bb\openone_\inp| \right),\quad
  \map{M} (\rho) = \Tr_\inp[(\openone_\out\otimes\transp{\rho})R_{\map M}],
\end{equation} 
where $\map{I}$ is the identity map and $\Tr_\inp$ denotes the partial trace on $\sH_\inp$ (see
\cite{DLclon} for the second equality in Eq. (\ref{R})). The transpose and the complex conjugate of a map are uniquely defined by through the relations
\begin{eqnarray}\label{taumap}
 ( \map I \otimes \map M^\tau) (|\openone_\out \kk \bb \openone_\out |) &=& R_{\map  M}\\ 
 \map M^*  &=&  (\map M^\dag )^\tau~, \label{starmap} 
\end{eqnarray}
where $\map M^\dag$ is the adjoint of the linear operator $\map M$ with respect to
the Hilbert-Schmidt scalar product, i.e. $\< B,  \map M (A)\>_{HS} = \< \map M^\dag
(B), A\>_{HS}$.  According to the above definitions, one has the useful relation
\begin{equation}\label{usefulchoi}
R_{\map B \map M \map A} = (\map B \otimes \map A^\tau) (R_{\map M})~,
\end{equation}
where $\map B$ is an arbitrary map from $\L{\sH_\out}$ to $\L{\sH_\out}$ and $\map A$ is an arbitrary map from $\L{\sH_\inp}$ to $\L{\sH_\inp}$.

It is easy to check that the linear map $\map{M}$ is CP if and only if
the {\bf CJ operator} $R_\map{M}$ is positive, and the correspondence
$\map M \leftrightarrow R_{\map M}$ is an isomorphism of positive
cones.  Moreover, $\map{M}$ is trace-non-increasing if and only if the
following dominance relation holds
\begin{equation}\label{TrR}
  \Tr_\out[R_\map{M}]\leq\openone_\inp,
\end{equation} 
the equal sign corresponding to the trace-preserving case of the
quantum channel.

Another convenient isomorphism is the one between linear maps $\map M$ from $\L{\sH_\inp}$ to $\L{\sH_\out}$ and linear operators $\check R_\map{M}$
from $\sH_\inp^{\otimes 2}$ to $\sH_\out^{\otimes 2}$ given by  
\begin{equation}\label{Rcheck}
\check R_{\map M} |A \kk = |\map M (A) \kk \qquad \forall A \in  \L{\sH_\inp}~, 
\end{equation}
such a definition depending on the two chosen basis $\{|c_m\>\}$
and $\{|e_n\>\}$ for $\sH_\out$ and $\sH_\inp$, respectively.  In this
case one has $\check R_{\map A \map B} = \check R_{\map A} \check
R_{\map B} $, namely the correspondence $\map M \leftrightarrow \check
R_{\map M}$ is an isomorphism of (finite dimensional) algebras.  The
correspondence $\map M \leftrightarrow \check R_{\map M}$ also induces
a one-to-one correpondence between $R_{\map M} \in \L{\sH_\out \otimes
  \sH_\inp}$ and $\check R_{\map M} \in \L{\sH_\inp^{\otimes 2},
  \sH_\out^{\otimes 2}}$:
\begin{equation}\label{ISO}
  \check{R}_\map{M}=\ISO(R_\map{M})~.
\end{equation}
Like $\check{R}_\map M$, the isomorphism $\ISO$ depends on the two
chosen basis $\{|c_m\>\}$ and $\{|e_n\>\}$ for $\sH_\out$ and
$\sH_\inp$.


\bigskip Every quantum operation $\map{M}$ can be written in a
(non-unique) {\bf Kraus form}
\begin{equation}\label{krausf}
  \map{M}(\rho)=\sum_i M_i\rho M^\dag_i~ \qquad \forall \rho \in \L{\sH_\inp}~.
\end{equation}
Any Kraus form is equivalent to a decomposition the CJ positive operator
$R_\map{M}$ into rank-one positive operators
\begin{equation}\label{Rsum}
  R_\map{M}=\sum_i
  |M_i\kk\bb M_i|~.
\end{equation}  
In particular, diagonalization of $R_\map{M}$ yields  the {\bf
  canonical Kraus form} $\map{M}(\rho)=\sum_i K_i\rho K_i^\dag, \
\Tr[K_i^\dag K_j] = \delta_{ij}\n{K_i}_2^2 $, where
$\n{X}_2:=\Tr[X^\dag X]^\frac12$ is the Hilbert-Schmidt norm.

For a map $\map M$ with Kraus form (\ref{krausf}), it is immediate to show that the maps $\map M^\dag$, $\map M^\tau$  and $\map M^*$ have the Kraus forms
\begin{equation}
\begin{array}{llll}
\map M^* (\rho) &=& \sum_i M_i^* \rho M_i^\tau  \qquad  &\forall \rho \in \L{\sH_\inp}\\
\map M^\dag (A) &=& \sum_i M_i^\dag A M_i \qquad &\forall A \in \L{\sH_\out}\\ 
\map M^\tau (A) &=& \sum_i M_i^\tau A M_i^* \qquad &\forall A \in \L{\sH_\out}~.
\end{array}
\end{equation}

Moreover, using Eqs.  (\ref{Rcheck}) and (\ref{kkprop}) the operator
$\check R_{\map M}=\ISO(R_\map{M})$ can be written in terms of any
Kraus form as follows
\begin{equation}
  \ISO(R_\map{M})=\sum_i M_i\otimes M_i^*.
\label{check}
\end{equation}
Different Kraus decompositions are all connected to a minimal
one---\eg the canonical---by an isometric matrix $W$ as follows
\begin{equation}\label{WM}
  M_i=\sum_jW_{ij}K_j,\quad \sum_{k}W_{ki}^*W_{kj}=\delta_{ij}.
\end{equation}
Therefore, the operator space $\Span\{M_i\}$ is independent of the choice of the
Kraus form, and is a function only of the map $\map{M}$. In the
following we will make use of the corresponding Hilbert space, which is spanned by the bipartite vectors $\{|M_i\kk\}$
\begin{equation}\label{HM}
  \sH_{\map M} =\Span\{|M_i\kk\}\equiv\Supp(R_\map{M}) \equiv \Rng (R_\map M)~,
\end{equation}
having used that the CJ operator $R_\map M$ is positive, whence support and range
coincide.  Note that generally $\sH_{\map M}$ can be a proper subspace
of $\sH_\out \otimes \sH_\inp$.

\subsection{Operator frames}
The Kraus operators $\{M_i\}$ are generally non-orthogonal, and not
even linearly independent. They are a so-called {\bf operator frame}
for the operator space $\Span\{M_i\}$, namely a (possibly infinite)
set of vectors such that the sum $\sum_i |M_i \kk \bb M_i|$ converges
to an operator $M \in \L{\sH_\out \otimes \sH_\inp}$, called the
{\bf frame operator}. Mor Kraus operators we have indeed
\begin{equation}
\MO:=\sum_i|M_i\kk\bb M_i|= R_{\map M}~.
\end{equation}

Any  vector $|A\kk$ in $\Supp(R_\map{M})$ can be expanded on the frame $\{|M_i\kk\}$, and the expansion  can be written in terms of another set of operators $\{N_i\}$
called {\bf dual of the frame} $\{M_i\}$ as  $   |A\kk=\sum_i |M_i\kk\bb N_i|A\kk$. Equivalently, we have the completeness relation
\begin{equation}\label{frameidentity}
 \sum_i |M_i\kk\bb N_i|= \openone_{\out}\otimes \openone_{\inp}~.
\end{equation}
A possible choice of dual, particularly relavant to our
purposes, is  the {\bf canonical dual} given by
\begin{equation}\label{canF}
  |\hat M_i\kk=\MO^{-1}|M_i\kk,
\end{equation}

The inverse $\MO^{-1}$ is actually defined on $\sH_{\map M}
=\Rng(R_{\map M})\equiv \Supp(R_{\map M})$, and on
$\sH_\out\otimes\sH_\inp$ it must be regarded as the Moore-Penrose
pseudoinverse with support on $\sH_{\map M}$.

\subsection{Minimal  dilation of a quantum operation}
For $\{M_i\}$ Kraus operators of the CP map $\map{M}$, the frame
operator $\MO$ is just the CJ operator $R_\map{M}\equiv\MO$ of the map
$\map{M}$, whence it is independent of the choice of the Kraus
operators $\{M_i\}$. Consider now the operator $V: \sH_\inp
\to \sH_\out \otimes \sH_{\map M}$ defined by 
\begin{equation}\label{isoV}
\begin{split} 
  V &: =\sum_i M_i\otimes \left ((\MO^\tau )^{ - \frac{1}{2} }|M_i^*\kk \right)\\ 
& = \sum_i M_i\otimes \left( (\MO^* )^{ -\frac{1}{2} }|M_i^*\kk \right) 
\end{split}
\end{equation}
having used $\MO^* = \MO^\tau$ since $\MO \ge 0$.
Note that  $V$ is independent of
the choice of the Kraus form: indeed, one has
\begin{equation}\label{Vconikraus}
  \begin{split}
  V &=\sum_{ijk }W_{ij} W_{ik}^* K_j\otimes \left( (R_\map{M}^*)^{-\frac{1}{2}  } |K_k^*\kk \right)\\
  &=\sum_j K_j \otimes \left( (R_\map{M}^*)^{-\frac{1}{2}} |K_j^*\kk \right)\\
  &=\sum_j K_j\otimes\frac{|K_j^*\kk}{\n{K_j}_2},
\end{split}
\end{equation}
having used Eq. (\ref{WM}) and the fact that each canonical Kraus
operator $K_i$ is eigenvector of $R_\map{M}$ with eigenvalue
$\n{K_i}_2^2$. Clearly the operator $V$ provides a dilation of the CP
map $\map M$, with $\sH_{\map M}$ playing the role of ancillary
Hilbert space:
\begin{equation}\label{stineqo}
  \map M(\rho) = \Tr_{\sH_{\map M}} [V \rho V^\dag]~.
\end{equation} 
For quantum operations $V$ is a contraction ($V^\dag V \le
\openone_\inp$), while for quantum channels $V$ is an isometry
($V^\dag V = \openone_\inp$). 

Among all possible dilations of $\map M$, the one given by $V$ in Eq.  (\ref{stineqo}) has minimum
ancilla dimension.  Indeed, for any operator $V' : \sH_\inp \to \sH_\out \otimes \sH_A$ such that 
$\map M (\rho) = \Tr_A [V'\rho V'^\dag]$, the map $\map M (\rho)$ has Kraus representation $\{M_i
:={}_A\<i|V\}$ where $\{|i\>_A \}$ is an orthonormal basis for $\sH_A$. Then, according to Eq.
(\ref{Rsum}) $\dim (\sH_A) \ge \rnk (R_{\map M}) =\dim (\sH_{\map M}) $. In other words,
$V$ is the {\bf minimal Stinespring dilation} of the CP map $\map M$ \cite{Stinespring}.  Any
non-minimal dilation $V'$ is connected to the minimal one via an isometry of ancillary spaces $Y :
\sH_{\map M} \to \sH_{A},\; \ Y^\dag Y = \openone_{\sH_{\map M}}$.  Indeed, using Eq. (\ref{WM}) one
has
\begin{equation}\label{V'fromV}
  V'=\sum_{ij}W_{ij}K_j\otimes|i\>=\sum_jK_j \otimes|\psi_j\> = (\openone_\out \otimes Y) V,
\end{equation}
where $\{|\psi_j\> \in \sH_A\}$ are the orthonormal vectors
$|\psi_j\>:=\sum_iW_{ij}|i\>$ and $Y$ is the isometry $Y:=
\frac{|\psi_j\>\<\!\< K_j^*|} {|\!| K_j |\!|_2 }$. The minimal
Stinespring dilation is unique up to local unitaries on the ancilla
Hilbert space $\sH_{\map M}$, namely if $V'$ is also  minimal, then $Y$ is a unitary from $\sH_{\map M}$ to $\sH_A$ \cite{Stinespring}.

\bigskip 

We now give a new compact formula for  the minimal Stinespring dilation of a CP map in terms of the Choi-Jamio\l kowski operator $R_{\map M}$: 

\begin{Prop}
  Let $R_{\map M} \in \L{\sH_\out, \sH_\inp}$ be the Choi-Jamio\l
  kowski operator associated to the CP map $\map M \in \CP {\sH_\inp,
    \sH_\out}$, and let $\sH_{\map M}$ be the Hilbert space $\sH_{\map M} = \Supp (R_\map M) = \Rng (R_{\map M})$. Then, a minimal dilation $V : \sH_\inp \to \sH_\out
  \otimes \sH_{\map M}$ is given by
\begin{equation}\label{isochoi}
  V  =  \left(\openone_{\out} \otimes (R^\tau_\map M)^{ \frac12} \right)  (|\openone_{\out}\>\!\> \otimes \openone_\inp) ~,
\end{equation}
or, alternatively, by
\begin{equation}\label{canonV}
  V=\left(\ISO(R_\map{M}^{\frac{1}{2}})\otimes\openone_\inp \right)(\openone_\inp\otimes|
  \openone_\inp\kk)~,
\end{equation}
$\ISO$ being the one-to-one correspondence defined in Eq. (\ref{ISO})
\end{Prop}  

\Proof It is simple to check that Eq. (\ref{isochoi}) provides a
dilation of $\map M$, which is clearly minimal since the ancilla space
is $\sH_{\map M} =\Supp (R_{\map M})$.  Indeed, using the inclusion
$\sH_\out \otimes \sH_{\map M} \subseteq \sH_\out^{\otimes 2} \otimes
\sH_{\inp}$ and Eqs.  (\ref{kkprop}) and (\ref{R}) one has for any
$\rho \in \St{\sH_\inp}$
\begin{equation}
\begin{split}
  \Tr_{\sH_{\map M}} [V\rho V^\dag ]  &= \Tr_{\out_2} \Tr_{\inp} [(\openone_{\out} \otimes R_{\map M}^\tau) (|\openone_\out\kk\bb \openone_\out |\otimes \rho) ]\\
 &= \Tr_{\out_2} \left [(\openone_{\out} \otimes \Tr_\inp[ (\openone_{\out} \otimes \rho ) R_{\map M}^\tau] ) ~(|\openone_\out\kk\bb \openone_\out | ) \right]\\
  & = \Tr_{\out_2}  \left[(\openone_{\out} \otimes \map M (\rho)^\tau) |\openone_\out \kk \bb \openone_\out|\right]\\
  &= \map M(\rho)~,
\end{split}
\end{equation} 
$\Tr_{\out_2}$ denoting partial trace over the second copy of
$\sH_{\out}$ in the tensor product $\sH_{\out}^{\otimes 2} \otimes
\sH_\inp$.  On the other hand, using the relation
$(R_\map{M})^s=\sum_j\n{K_j}^{2(s-1)}|K_j\kk\bb K_j|$ along with
Eq.~\eqref{check} we get $\ISO( R_{\map M}^{\frac 1 2}) = \sum_j
\|K_j\|^{-\frac 1 2} K_j \otimes K_j^*$. Substituting $\ISO (R_{\map
  M}^{\frac 1 2})$ into Eq. (\ref{canonV}) and comparing with Eq.
(\ref{Vconikraus}) we obtain that the definition of $V$ in Eqs.
(\ref{canonV}) and (\ref{isoV}) actually coincide.  Finally, direct
calculation  shows the coincidence of definitions of $V$ in (\ref{isochoi}) and (\ref{canonV}). \qed


\subsection{POVMs and Quantum Instruments}
The statistics of a quantum measurement is described by a {\bf
  measurable space} $\Omega$ with a {\bf $\sigma-$algebra}
$\Sigma_\Omega$ of {\bf events}, and a {\bf probability measure} $p$
on $(\Omega,\Sigma_\Omega)$. In Quantum Mechanics the probability
measure in terms of the quantum state $\rho$ is given by the {\bf Born
  rule}
\begin{equation}
  \forall B\in\Sigma_\Omega,\quad p(B)=\Tr[\rho P_B],
\end{equation}
where $P$ is a {\bf positive operator valued measure (POVM)},
namely a map from events $B\in\Sigma_\Omega$ to positive operators
$P_B\geq 0$ on $\sH$, satisfying the requirements
\begin{align}\label{NormPovm}
  &P_\Omega=\openone\quad&\text{(normalization)}\\
  &P_{\left ( \bigcup_{i=1}^{\infty} B_i \right)}=\sum_{i=1}^{\infty}
  P_{B_i}\quad \forall \{B_i\} :\; B_i \cap B_j = \emptyset \; \forall
  i \not = j,\quad &\text{($\sigma$-additivity)},
\end{align}
where the series converges in the weak operator topology.

A complete description of a measurement in a cascade of different measurements performed on the same
system must also provide the {\bf conditional state} associated to any possible event.  In Quantum
Mechanics this is given by the notion of "quantum instrument" \cite{Ozawacontextens}, in which
each event $B\in\Sigma_\Omega$ corresponds to a quantum operation 
$\map{Z}_B\in\CP{H_\inp,H_\out}$. More precisely, we have
\begin{Def}
  A map $\map{Z}: \Sigma_\Omega\to\CP{H_\inp,H_\out}$ is a
  {\bf quantum instrument} if it satisfies the properties
\begin{eqnarray}
  \label{NormInst}  &&\Tr[\map{Z}_\Omega (\rho)] = \Tr[\rho]  \qquad \qquad  \quad \forall \rho \in \St{\sH_{\inp}}\\
&&\map{Z}_{\left ( \bigcup_{i=1}^{\infty} B_i \right)}= \sum_{i=1}^{\infty} \map{Z}_{B_i}  \qquad  \forall \{B_i\} : \quad B_i \cap B_j = \emptyset \quad \forall i \not = j~.
\end{eqnarray}
\end{Def} 

Using the CJ isomorphism (\ref{R}), any instrument $\map{Z}$ can be
associated in a one-to-one fashion with a positive operator valued
measure $Z$, which we call {\bf Choi-Jamio\l kowski measure (CJM) of the instrument}, given by 
\begin{equation}\label{RInst}
  Z_B := R_{{\map Z}_B} =\map{Z}_B \otimes\map{I}\left( |\openone_\inp\kk\bb\openone_\inp| \right) \qquad \forall B \in
  \Sigma_\Omega.  
\end{equation} 
Differently from usual POVMs, for which the normalization is given by Eq. (\ref{NormPovm}) the measure $Z$ has the normalization condition
\begin{equation}
\Tr_\out[Z_\Omega] = \openone_\inp~,
\end{equation}
$\Tr_{\out} [\cdot]$ denoting partial trace over $\sH_{\out}$.  

The POVM $P$ giving the probability of the event
$B\in\Sigma_\Omega$ for state $\rho\in\St{H_\inp}$ can be written in
terms of the CJM $Z$ using the isomorphism (\ref{R}) as follows
\begin{equation}
\begin{split}
  \Tr[P_B\rho]&=\Tr[\map{Z}_B(\rho)]=\Tr[(\openone_\out\otimes\transp{\rho})Z_B] =\Tr[\rho \Tr_\out[Z^\tau_B]]
\end{split}
\end{equation}
whence
\begin{equation}
  P_B=\Tr_\out[Z_B^\tau]\qquad \forall B\in\Sigma_\Omega. 
\end{equation}

In finite dimensions, the correspondence between instruments and CJMs
allows one to simply prove the existence of an instrument density
w.r.t. a suitable scalar measure
\cite{Holevo:1998p3165,Ozawa:1985p370}.
\begin{Prop}\label{Theo:InstDens}
  Any instrument $\map{Z} : \Sigma_\Omega\to\CP{H_\inp,H_\out}$ in
  finite dimensions can be written as
\begin{equation}
  \map{Z}_B = \int_B \mu (\d\omega)\map{S}_\omega  
\end{equation}
where $\mu (\d \omega)$ is the finite measure defined by $\mu
(B)\doteq\Tr[Z_B]$ $\forall B \in \Sigma_\Omega$, and the density
$\map{S}_\omega $ is a CP-map valued function, uniquely defined
$\mu-$almost everywhere.
\end{Prop}

\Proof Let $\{|k\>\}$ be an orthonormal basis for $\sH_\out \otimes
\sH_\inp$ and define the complex measures $\mu_{kl}, \ \mu_{kl} (B)
:=\<k| Z_B|l\>$. Due to positivity, one has $|\mu_{kl}(B)| \le \sqrt{
  \<k| Z_B |k\>\<l|Z_B |l\>} \le \Tr [ Z_B]=\mu(B)$, i.e.  all
measures $\mu_{kl}$ are absolutely continuous w.r.t.  $\mu$.
Therefore, any measure $\mu_{kl}$ admits a density
$\sigma_{kl}(\omega)$ w.r.t. $\mu$. We then
have
\begin{equation}
  Z_{B} = \sum_{k,l} \mu_{kl}(B) |k\>\<l|  =  \sum_{k,l} \int_{B} \mu(\d \omega) \sigma_{kl} (\omega) ~|k\>\<l| = \int_B
  \mu(\d \omega) ~S_\omega~,
\end{equation}
having defined $S_\omega:= \sum_{k,l} \sigma_{kl} (\omega)|k\>\<l|$.
Since $S_\omega $ is the density of the positive operator valued
measure $Z_B$ w.r.t. the scalar measure $\mu$, it is positive and
uniquely defined $\mu$-a.e.  The instrument density $\map{S}_\omega $
is then obtained by the relation $\map{S}_\omega (\rho) =
\Tr_\inp[(\openone_\out\otimes\transp{\rho})S_\omega]$.\qed

\section{Dilations of Quantum Instruments}\label{sec:gendil}
We are now in position to prove a dilation theorem for instruments
with generally continuous outcome space in finite dimensions.  A {\bf
  dilation of a quantum instrument} $\map Z: \Sigma_\Omega \to
\CP{H_\inp,H_\out}$ is a triple $(\sH_A, V, Q)$, where $\sH_A$ is an
ancillary Hilbert space, $V: \sH_\inp \to \sH_\out \otimes \sH_A$ is
an isometry, and $Q: \Sigma_\Omega \to \L{\sH_A}$ is a POVM on the
ancilla, such that
\begin{equation}\label{defdilation}
  \map Z_B (\rho) = \Tr_{\sH_A} [ V\rho V^\dag (\openone_\out \otimes Q_B)] \qquad \forall B \in \Sigma_\Omega~.
\end{equation}
The triple $(\sH_A, V, Q)$ represents an {\bf indirect
  measurement scheme} where the input system $\sH_\inp$ evolves
through the isometry $V$, producing the output $\sH_\out$ and the
ancilla $\sH_A$, which then undergoes a POVM measurement $Q$
with the same outcome space as the instrument.  
\begin{Theo}\label{t:dilations} 
  Let $\map{Z}: \Sigma_\Omega\to\CP{H_\inp,H_\out}$ be an instrument
  with outcome space $\Omega$, and $Z: \Sigma_\Omega \to \L
  {\sH_\out\otimes \sH_\inp}$ be the associated CJM. A minimal
  dilation of the instrument is given by the triple $(\sH_A, V, Q)$
  where the ancillary Hilbert space $\sH_A$ is isomorphic to
  $\sH_{\map Z}:=\Supp(Z_\Omega)= \Rng(Z_\Omega)$, $V: \sH_\inp \to \sH_\out \otimes
  \sH_A$ is the isometry
\begin{equation}
V := \left(\openone_\out \otimes  (Z_\Omega^\tau)^{\frac 1 2} \right) (|\openone_\out \kk \otimes \openone_\inp)~,
\end{equation}
 and
$Q$ is the POVM on $\sH_A$ given by  
\begin{equation}\label{QB}
  Q_B:=\left(Z_\Omega^{-\frac{1}{2}}Z_BZ_\Omega^{-\frac{1}{2}}\right)^\tau \qquad \forall B\in\Sigma_\Omega  .
\end{equation}
\end{Theo}
\Proof According to Eq. (\ref{isochoi}), $V$ is the minimal Stinespring isometry of the channel
$\map Z_\Omega$. on the other hand, $Q$ is clearly a POVM on $\sH_A$, since $Q_B \ge 0 \ \forall B
\in \Sigma_\Omega$ and $Q_\Omega = \openone_{\sH_A}$.  Moreover, exploiting the inclusion $\sH_\out
\otimes \sH_A\subseteq \sH_\out^{\otimes 2} \otimes \sH_\inp$ we have
\begin{equation}
\begin{split}
\Tr_{\sH_A} [V \rho V^\dag (\openone_\out \otimes Q_B)] &= \Tr_{\out_2} \Tr_{\inp} [(|\openone_\out \kk \bb \openone_\out| \otimes \rho) (\openone_\out \otimes Z_B^\tau)]\\
&= \Tr_{\out_2} [(\openone_\out \otimes \map Z_B (\rho)^\tau ) |\openone_\out \kk \bb \openone_\out|] =\map Z_B (\rho)~, 
\end{split}
\end{equation}
thus proving that $(\sH_A, V, Q)$ is actually a dilation of the
instrument $\map Z$.
Finally, the dilation has minimal ancilla dimension. Indeed, for any
dilation $(\sH_{A'}, V',Q')$ of the instrument $\map Z$, we have a
dilation of the channel $\map Z_\Omega$, given by $\map{Z}_{\Omega}
(\rho)=\Tr_{\sH_{A'}}[V'\rho V'^\dag ]$. Since $V$ is the minimal
Stinespring isometry of the channel $\map Z_\Omega$, one necessarily
has $\dim (\sH_{A'}) \ge \dim (\Supp(Z_{\Omega})) \equiv \dim
(\sH_{\map Z})$.  \qed

Any other dilation of the instrument $\map{Z}$ arises from some
non-minimal isometry $V': \sH_\inp \to \sH_\out \otimes \sH_A$, which
is necessarily of the form $V' = (\openone_\out \otimes Y) V$,
where $Y: \sH_{\map Z} \to \sH_A$ is an isometry of ancilla spaces.
Substituting the form of $V'$ in Eq. (\ref{defdilation}) we then obtain 
\begin{equation}\label{compression}
  Q_B=Y^\dag Q_B'Y \qquad \forall B \in \Sigma_\Omega~.
\end{equation}
Since $Y$ can be viewed as an isometric embedding of $\sH_{\map Z}$
into $\sH_A$, the above equation means that $Q$ is the projection of
$Q'$ on the support of $Y$.  This is indeed the case of the
non-minimal dilation provided by Ozawa's dilation theorem
\cite{Ozawacontextens}, where $Q'$ is a projection valued measure
(PVM) on the infinite dimensional ancilla space $\sH_A$.  According to
Eq.  (\ref{compression}), $Q'$ is then a Naimark dilation
\cite{NaimarkMA40} of the minimal POVM $Q$ provided in our theorem.

\section{Instruments generated by operator
  frames}\label{sec:frame-orbit}

We now introduce the definition of {\bf frame-orbit instruments},
which will play an important role in the construction of feed-forward
realization schemes and generalized teleportation schemes.

Fix a finite measure $\mu (\d \omega)$ on the outcome space $\Omega$,
a measurable family of quantum operations $\oA_\Omega :=\{\map
A_\omega\}_{\omega \in \Omega}$
, and a measurable family of quantum channels $\sB_\Omega := \{\map
B_{\omega}\}_{\omega \in \Omega}$. Then we have the following:

\begin{Def}[Frame-orbit instruments]\label{Def:CovInst}
  An instrument $\map{Z}: \Sigma_\Omega\to\CP{H_\inp,H_\out}$ is a
  \emph{frame-orbit instrument w.r.t. $(\mu, \oA_\Omega,\sB_\Omega)$}
  if $\map Z$ admits a density w.r.t. $\mu$ and the density has the
  form
  \begin{equation}\label{frameorbit}
    \map S_\omega = \map B_\omega \map S_0 \map A^\dag_\omega \qquad \mu-\forall \omega \in \Omega~,
  \end{equation}
  where $\map S_0$ is a fixed CP map.
  In the case $\map B_\omega \equiv \map I_\out \ \forall \omega \in
  \Omega$ we say that $\map Z$ is a frame-orbit instrument w.r.t.
  $(\mu,\oA_\Omega)$.
\end{Def}

According to proposition \ref{Theo:InstDens}, any instrument $\map Z$
can be trivially viewed as a frame-orbit instrument by taking $ \mu(B)
:= \Tr[Z_B]$, $\map A_\omega\equiv\map S_\omega$, $\map S_0\equiv\map
I$ and $\map B_\omega\equiv \map I$. However, a given instrument can
be a frame-orbit instrument w.r.t. several different triples $(\mu,
\oA_\Omega, \sB_\Omega)$, and, on the contrary, once a triple
$(\mu,\map A_\Omega,\map B_\Omega)$ has been fixed, not all
instruments are frame-orbit instruments w.r.t. that triple.


From now on we will restrict our attention to the case where the elements
of $\map A_\Omega$ are {\bf single-Kraus operations} $\map A_\omega(\cdot):=
A_\omega\cdot A_\omega^\dag$. The generalization of all results to the
case $\map A_\omega (\cdot)= \sum_{k=1}^{d_\inp^2} A_{\omega,k} \cdot A_{\omega,k}^\dag$ is straightforward, as it only
consists in replacing the index $\omega$ by the couple $(\omega,k)$,
$\int_\Omega\mu(\d\omega)$ by $\int_\Omega\mu(\d\omega)
\sum_{k=1}^{d_\inp^2}$ and taking $\map B_{(\omega,k)}:=\map B_\omega$
and $\map S_\omega:=\sum_{k=1}^{d_\inp^2}\map S_{(\omega,k)}$.

\begin{lemma}
  Let $\map Z : \Sigma_\Omega\to\CP{H_\inp,H_\out}$ be a frame-orbit
  instrument w.r.t. $(\mu, \oA_\Omega,\sB_\Omega)$ with density $\map
  S_\omega = \map B_\omega \map S_0 \map A_\omega^\dag$, $\map S_0
  (\cdot) = \sum_{i=1}^r S_i \cdot S_i^\dag$ be a Kraus form for $\map
  S_0$, and $\xi := \sum_{i=0}^r S_i^\dag S_i \in \L {\sH_\inp}$.
  Then,
\begin{equation}\label{frameorbitnorm}
  \int_\Omega \mu (\d \omega) A_\omega \xi A_\omega^\dag = \Tr_{\inp_2}[ (\openone_{\inp} \otimes \transp \xi ) A] = \openone_\inp ~,
\end{equation}
where $A \in \L{\sH_{\inp}^{\otimes 2}}$ is the frame operator
\begin{equation}\label{frameopa}
  A = \int_\Omega \mu (\d \omega) |A_\omega \kk \bb A_\omega|~,
\end{equation}
and $\Tr_{\inp_2}$ denotes partial trace over the second copy of $\sH_\inp$ in the tensor $\sH_\inp^{\otimes 2}$.
Viceversa, for any positive operator $\xi \in \L {\sH_\inp}$
satisfying Eq. (\ref{frameorbitnorm}) there exists a frame-orbit
instrument w.r.t. $(\mu, \oA_\Omega, \sB_\Omega)$.
\end{lemma}

\proof For the normalization of the instrument $\map Z$, $\map
Z_\Omega$ must be trace-preserving, and we have $\Tr[\map Z_\Omega
(\rho)]= \int_\Omega \mu (\d \omega) \Tr[ A_\omega \xi A_{\omega}^\dag
\rho] = \Tr[\rho] \ \forall \rho \in \St {\sH_\inp}$, whence Eq.
(\ref{frameorbitnorm}).  Viceversa, for any $\xi \ge 0$ satisfying Eq.
(\ref{frameorbitnorm}) we can define $\map S_0 (\cdot) := {\xi}^\frac12
\cdot {\xi}^\frac12$, so that $\map S_\omega := \map B_\omega \map S_0
\map A_\omega^\dag$ is the density of a normalized frame-orbit
instrument. \qed

\medskip 

In particular, whenever the elements of $\oA_\Omega$ are all proportional to unitary channels, the
class of frame-orbit instruments w.r.t. $(\mu, \oA_\Omega,\sB_\Omega)$ is nonempty, as one can
choose e.g. $\xi = \kappa \openone_\inp$ with suitable normalization constant $\kappa >0$. As we
will see in the following Section, this includes the case of group-covariant instruments. Similarly,
if $\sA_\Omega:=\{A_\omega\}$ is a tight operator frame, namely $A= \openone_\inp^{\otimes 2}$, by
definition Eq. (\ref{frameorbitnorm}) holds for any $\xi \in \L{\sH_\inp}$.  Notice that, however,
the operators in $\sA_\Omega$ do not need to be unitary in general, nor $\sA_\Omega$ needs to be a
tight frame, since it is enough that Eq.  (\ref{frameorbitnorm}) holds for a single operator $0\le
\xi \in \L {\sH_\inp}$.
\subsection{Canonically associated POVMs and their densities}
According to the previous section, there are two POVMs $P$ and $Q$
that are canonically associated to the instrument $\map{Z}$. The POVM
$P$ gives the probability distribution of the instrument for each
event and each state, whereas the POVM $Q$ allows to express the
minimal dilation of the instrument via the minimal isometry
$V=(\openone_\out \otimes {Z_\Omega^{\tau \frac12}}) (|\openone_\out\kk
\otimes \openone_\inp)$.  Both $P$ and $Q$ can be written in terms of
the CJM
$Z_B=\map{Z}_B\otimes\map{I}(|\openone_\inp\kk\bb\openone_\inp|)$ of
the instrument (see Eq. (\ref{RInst})) as
\begin{equation}\label{PQ}
  P_B=\Tr_\out [Z_B^\tau],\quad Q_B=\left(Z_\Omega^{-\frac{1}{2}}Z_BZ_\Omega^{-\frac{1}{2}}\right)^\tau.
\end{equation}
Obviously, since the instrument $\map Z$ admits a density w.r.t.  $\mu$, also the CJM $Z$ will admit
a density w.r.t. $\mu$, given by
\begin{equation}\label{CJMdensity}
  S_\omega : = (\map S_\omega \otimes \map I) (|\openone_{\inp}\kk\bb \openone_{\inp}|)~,
\end{equation}
which is positive and uniquely defined $\mu$-almost everywhere.  From
Eqs. (\ref{frameorbit}), (\ref{usefulchoi}),and (\ref{starmap}) it is also clear that the density $S_\omega$
has the form
\begin{equation}\label{CJdensity}
  S_\omega = (\map B_\omega \otimes \map A_\omega^*) ( S_0)~,  
\end{equation}
having defined $S_0 := (\map S_0 \otimes \map I) (|\openone_{\inp}\kk\bb
\openone_{\inp}|) = \sum_{i=1}^r |S_i \kk \bb S_i|$. Finally, from Eq.
(\ref{PQ}) it follows that the POVMs $P$ and $Q$ admit densities
w.r.t. $\mu$, $\xi_\omega$ and $\zeta_\omega$, respectively, given by
\begin{equation}\label{otherdensities}
  \begin{split}
    \xi_\omega &=  A_\omega \xi  A_\omega^\dag  \qquad \xi := \Tr_{\out} [S_0^\tau]= \sum_{i=1}^r S_i^\dag S_i \\
    \zeta_\omega &= \left( Z_{\Omega}^{-\frac12} S_\omega Z_\Omega^{-\frac12}
    \right)^\tau~.
  \end{split}
\end{equation}

\subsection{Feed-forward realization of frame-orbit
  instruments}\label{s:feed}
The realization of frame-orbit instruments w.r.t. $(\mu,
\oA_\Omega,\sB_\Omega)$ can be always reduced to the realization of
frame-orbit instruments w.r.t. $(\mu, \oA_\Omega)$, combined with a
feed-forward classical communication to implement the conditional
channel $\map B_\omega$. Indeed, according to Eq. (\ref{frameorbit}),
every frame-orbit instrument $\map Z$ w.r.t.
$(\mu,\oA_\Omega,\sB_\Omega)$ is equivalent to the frame-orbit
instrument $\map T$ w.r.t.  $(\mu,\oA_\Omega)$ given by
\begin{equation}\label{T}
  \map T_B = \int_{B} \mu(\d \omega)  \map S_0 \map A_\omega^\dag 
\end{equation}
followed by the channel $\map{B}_{\omega}$ depending on the outcome
$\omega$. Notice that $\map T$ is a normalized instrument, since $\map
T$ and $\map Z$ have the same normalization in Eq.
(\ref{frameorbitnorm}).

According to Eq. (\ref{CJdensity}) the CJ operator $T_\Omega = \map
T_\Omega \otimes \map I (|\openone_\inp \kk \bb \openone_\inp|)$ is
then given by
\begin{equation}\label{tomega}
  \begin{split} 
    T_\Omega &= \int_\Omega \mu (\d \omega)  (\map I \otimes \map A_\omega^*) ( S_0)\\
&=  (\map S_0 \otimes \map I) \left(\int_\Omega \mu (\d \omega) (\map I \otimes \map A_\omega^*) ( |\openone_\inp \kk \bb \openone_\inp|) \right)\\
    &= (\map S_0\otimes \map I)(E A^*E)
  \end{split}
\end{equation}
with $\map S_0 \in \CP{\sH_{\inp_1},\sH_{\out}}$, and $A \in \L
{\sH_{\inp}^{\otimes 2}}$ being the frame operator in Eq.
(\ref{frameopa}), and $E$ denoting the unitary swap
between the two copies of $\sH_{\inp}$ in the tensor $\sH_\inp^{\otimes 2}$.
Combining the feed-forward scheme with the minimal dilation of the instrument $\map T$ we obtain the following
\begin{Cor}\label{Cor:feed-for}
  Let $\map Z \in \CP {\sH_\inp,\sH_\out}$ be a frame-orbit instrument
  w.r.t. $(\mu, \oA_\Omega, \sB_\Omega)$ with density $\map S_\omega =
  \map B_\omega \map S_0 \map A_\omega^\dag$, and let $T_\Omega\map
  \in \L{\sH_\out \otimes \sH_\inp}$ be the CJ operator defined in
  Eq. (\ref{tomega}). Then, the instrument $\map Z$ has the
  minimal feed-forward realization
  \begin{equation}
    \map S_\omega (\rho)= \map B_\omega \left( \Tr_{\sH_A} \left[   V \rho V^\dag \, (\openone_\out \otimes \zeta_\omega)  \right] \right)~,
  \end{equation}
  where $V $ is the isometry $V := \left( \openone_\out \otimes
    (T_{\Omega}^\tau)^{ \frac12}\right) ( |\openone_\out\kk \otimes
\openone_\inp)$, and $\zeta_\omega$ is the POVM density $\zeta_\omega
:= \left(T_\Omega^{-\frac12} (\openone_\out \otimes A_\omega^*) S_0
  (\openone_\out \otimes A^\tau_\omega) T_\Omega^{-\frac12}
\right)^\tau$.
\end{Cor}

Feed-forward schemes have recently attracted a remarkable interest in
quantum optics, and have been experimentally demonstrated in several
applications, such as signal amplification \cite{feedforampli},
coherent state cloning \cite{feedforclon}, minimum-disturbance
estimation \cite{feedforest}, and squeezed state purification
\cite{feedforsqeez}.  In the finite-dimensional case, frame-orbit
instruments provide the most general mathematical framework in which
similar realization schemes can be searched.

\section{Group-covariant instruments with transitive outcome space}\label{sec:covinst}
A particular case of frame-orbit instruments is that of covariant
instruments whose outcome space $\Omega$ is a transitive $\grp
G$-space. Given a group $\grp G$, we will denote by $\sU
_\grp{G}:=\{U_g \in \L{\sH_\inp}\}_{g\in\grp{G}}$ by $\sV_{\grp G}:=
\{ V_g \in \L{\sH_\out}\}_{g \in \grp G}$ two unitary representations
of $\grp G$, and by $\mU_{\grp G}:=\{\mU_g \in \CP{\sH_\inp,
  \sH_\inp}\}_{g\in\grp{G}}$, $\mV_{\grp G}:=\{\mV_g \in \CP{\sH_\out,
  \sH_\out}\}_{g\in\grp{G}}$ the corresponding sets of automorphisms
$\mU_g(\cdot):=U_g\cdot U_g^\dag$, $\mV_g(\cdot):=V_g\cdot V_g^\dag$.
\begin{Def}[Group-covariant instruments]\label{Def:UCovInst}
  Given a topological group $\grp{G}$ acting on $\Omega$, and two
  continuous unitary (generally projective) representations $\sU
  _\grp{G}$ and $\sV_\grp{G}$ on the Hilbert spaces $\sH_\inp$ and
  $\sH_\out$, respectively, we say that the instrument $\map{Z}:
  \Sigma_\Omega\to\CP{H_\inp,H_\out}$ is {\em group-covariant} w.r.t.
  $(\grp{G},\mU_\grp{G},\mV_\grp{G})$ when one has
  \begin{equation}\label{GCovInst}
    \map{Z}_{B} \circ \map U_g(\rho) = \map V_g \circ \map{Z}_{g^{-1} (B)} (\rho)
    \qquad\forall\rho\in\St {H_\inp},\;\forall B \in \Sigma_\Omega,\;\forall g\in\grp{G}~,
  \end{equation}
  with $g^{-1} (B) :=\{\omega \in \Omega ~|~ g \omega \in B\}$.
\end{Def}

In the case of transitive group action on the outcome space $\Omega$,
for any point $\omega_0 \in \Omega$ one has $\Omega = \grp G
\omega_0$, and the outcome space $\Omega$ can be identified with the
space of left-cosets $\Omega \equiv \grp G / \grp G_0$ with respect to
the stability group $\grp G_0:=\{h \in \grp G~|~ h \omega_0 =
\omega_0\}$. Denote by $\pi$ the projection map $\pi : \grp G \to
\Omega , g \mapsto \pi(g) = g\omega_0$ and by $\mu$ the invariant
measure on $\Omega$ given by $\mu (B) = \int_{\pi^{-1} (B)} \d g$,
where $\d g$ is the normalized Haar measure over $\grp G$.  Under this
hypothesis and notation the following structure theorem holds
\cite{Davies,Holevo82}:
\begin{Theo}
  Let $\grp G$ be a compact group, $\grp G_0$ be a closed subgroup,
  and $\map Z : \Sigma_\Omega\to\CP{H_\inp,H_\out}$ be a covariant
  instrument w.r.t.  $(\grp G, \mU_{\grp G}, \mV_{\grp G})$ with
  outcome space $\Omega = \grp G/\grp G_0$. Then $\map Z$ admits a
  density w.r.t.  $\mu$ of the form
  \begin{equation}\label{covdens}
    \map S_{\pi (g)} = \map V_g \map S_0 \map U^\dag_g \qquad \forall g \in \grp G~,  
  \end{equation}
  where $\map S_0$ is a CP map, $\map V_g (\cdot):= V_g \cdot V_g^\dag
  $, and $\map U_g^\dag (\cdot) := U_g^\dag \cdot U_g$.
\end{Theo}
From now on we will confine our attention to compact groups $\grp G$. Since the group-action is transitive, Eq. (\ref{covdens}) defines the
density $\map S_\omega$ for any $\omega \in \Omega$. Indeed one can
take any measurable section $\sigma : \Omega \to \grp G, \omega
\mapsto \sigma (\omega), \, \pi(\sigma (\omega)) = \omega$ and declare
\begin{equation}\label{covdens2}
  \map S_\omega = \map V_{\sigma (\omega)} \map S_0 \map U_{\sigma (\omega)}^\dag~. 
\end{equation} 
The above equation clearly characterizes any covariant instrument with
$\Omega = \grp G/\grp G_0$ as a frame-orbit instrument w.r.t.
$(\mu, \mU_{\sigma(\Omega)}, \mV_{\sigma (\Omega)})$. In addition, Eq.
(\ref{covdens}) implies the invariance condition
\begin{equation}
  \map S_0 = \map V_h \map S_0 \map U_h^\dag \qquad \forall h \in \grp G_0~,
\end{equation}
which in terms of CJ operators becomes the commutation relation
\begin{equation} 
[S_0, V_h \otimes U_h^*]=0 \qquad h \in \grp G_0~.
\end{equation}

\subsection{Covariant POVMs and dilation of covariant instruments}

\subsubsection{Minimal dilation} 
Let $\map Z$ be an instrument  with outcome space $\Omega = \grp G/\grp G_0$ covariant w.r.t. $(\grp G, \mU_\grp G,
\mV_\grp G)$.  From Eqs.
(\ref{CJdensity}) and (\ref{covdens2}), the CJM density is
\begin{equation}
  S_\omega =  (\map V_{\sigma (\omega)} \otimes \map U^*_{\sigma (\omega)} ) (S_0)~.
\end{equation} 
Exploiting the Mackey-Bruhat identity, which sets up an isomorphism
between $\grp G$ equipped with the Haar measure $\d g$ and $\Omega
\times \grp G_0$ equipped with the product measure $\mu(\d \omega)
\times \nu(\d h)$, $\nu(\d h)$ normalized Haar measure over $\grp G_0$
(see e.g. \cite{Folland1995}), we obtain
\begin{equation}
  \begin{split}
    Z_\Omega & = \int_\Omega \mu (\d \omega) ~ S_\omega = \int_{\Omega
      \times \grp G_0} \mu (\d \omega) \nu(\d h) ~ (\map
    V_{\sigma(\omega)} \otimes \map U^*_{\sigma (\omega)} ) (S_0) = \\
    & =\int_{\grp G} \d g ~ (\map V_g \otimes \map U^*_g ) (S_0) ~.
  \end{split}
\end{equation}
As a consequence,we have $[Z_\Omega, V_g \otimes U^*_g]=0 \ \forall g
\in \grp G$ and the density $\zeta_\omega$ in Eq.
(\ref{otherdensities}) is given by
\begin{equation}
  \zeta_\omega = \left( V^*_{\sigma (\omega)} \otimes U_{\sigma (\omega)} \right) \Xi  \left( V^*_{\sigma (\omega)} \otimes U_{\sigma (\omega)} \right)^\dag \qquad \Xi:= \left( Z_\Omega^{-\frac12}  S_0 Z_\Omega^{-\frac12} \right)^\tau~,
\end{equation}
with $\Xi$ satisfying the commutation relation
\begin{equation} 
[\Xi, V^*_h \otimes U_h ] = 0 \qquad \forall h \in
  \grp G_0~.
\end{equation}
This shows that POVM $Q_B = \int_B \mu (\d \omega) ~ \zeta_\omega $
used in the minimal dilation of Theorem \ref{t:dilations} is a {\bf
  covariant POVM} with outcome space $\Omega$ \cite{Holevo82}, i.e.
\begin{equation}
  Q_{g (B)} = (\map V^*_g \otimes \map U_g) (Q_B) \qquad \forall B \in \Sigma_\Omega \ .
\end{equation} 
Furthermore, using the relation $[Z^\tau_\Omega, V_g^*\otimes
U_g]=[Z^*_\Omega , V_g^*\otimes U_g]=0 \ \forall g\in \grp G$, it is
immediate to see that the minimal isometry $V = \left(\openone_\out \otimes
 (Z_\Omega^\tau)^{\frac 12}\right) (|\openone_\out \kk \otimes \openone_\inp)$
intertwines the two representations $\sV_\grp G \otimes \sV^*_\grp G
\otimes \sU_\grp G$ and $\sU_\grp G$, namely
\begin{equation}
  ( V_g \otimes V_g^* \otimes U_g) ~ V = V ~ U_g \qquad \forall g \in \grp G~.
\end{equation}

\subsubsection{Minimal feed-forward realization and generalized teleportation schemes}

A minimal feed-forward realization can be obtained by using Corollary
\ref{Cor:feed-for}, in terms of the instrument $\map T_B := \int_B
\mu(\d \omega)~ \map S_0 \map U_{\sigma(\omega)}^\dag$.  In this case
we have
\begin{equation}\label{Tcov}
  T_\Omega = \int_\grp G \d g (\map I_\out \otimes \map U^*_g) (S_0)
\end{equation} 
whence $[T_\Omega, \openone_\out \otimes U^*_g ]=0 \ \forall g\in \grp
G$. As a consequence, the POVM density $\zeta_\omega$ is now given by
\begin{equation}
  \zeta_\omega  = \left( I\otimes U_{\sigma (\omega)} \right) \Xi  \left( I \otimes U_{\sigma (\omega)} \right)^\dag \qquad \Xi:= \left( T_\Omega^{-\frac12}  S_0 T_\Omega^{-\frac12} \right)^\tau 
  \ .
\end{equation}
Notice that in this case $\zeta_\omega$ is not a covariant POVM
density, since the relation $ [\Xi, \openone_\out \otimes U_h] =0 \
\forall h\in \grp G_0 $ does not necessarily hold. The minimal
isometry $V$ is now given by $V = \left(\openone_\out \otimes
(T_\Omega^\tau)^{\frac12}\right)  (|\openone_\out\kk \otimes \openone_\inp)$ and enjoys
the property
\begin{equation}
  ( \openone_\out^{\otimes 2} \otimes U_g) ~ V = V U_g  \qquad \forall g \in \grp G~.
\end{equation}

For irreducible $\sU_\grp G$ the above equation yields $V = |\Psi\kk
\otimes \openone_\inp$ for some $|\Psi\kk \in \sH_\out^{\otimes 2},
\n{\Psi}_2 =1$, namely the isometry $V$ is just the extension with
some pure state.  Precisely, computing the average in Eq.
(\ref{Tcov}) we have $T_\Omega = d_\inp^{-1}\Tr_\inp[ S_0] \otimes
\openone_{\inp}$, whence
\begin{equation}
  V = |\sigma^\frac12\kk \otimes \openone_\inp \qquad \sigma := d_\inp^{-1}\Tr_\inp [S_0]  = d_\inp^{-1} \sum_{i=1}^r S_i S_i^\dag ~.
\end{equation}
The feed-forward realization then becomes a {\bf generalized teleportation
  scheme} where $|\sigma^\frac12 \kk$ plays the role of entangled
resource, the joint measurement $\zeta_\omega$ is performed by the
sender on the input system and on half of the entangled state, the
outcome $\omega$ is classically transmitted, and the conditional
operation $\map V_{\sigma (\omega)}$ is performed at the receiver's
end.  The discussion on generalized teleportation schemes will be extended in
Section \ref{tele} to the case of frame-orbit instruments
generated by tight operator frames.

\subsubsection{Non minimal feed-forward dilations}
Using group theory it is easy to construct non-minimal dilations of
group-covariant instruments.  Let us decompose $\sH_\inp$ and
$\sU_\grp G$ as
\begin{equation}\label{RepDecomp}
  \sH_\inp=\bigoplus_{\mu \in \set S}\sH_\mu\otimes\Cmplx^{m_\mu}, \qquad 
  U_g =\bigoplus_{\mu \in \set S} U_g^\mu\otimes \openone_{m_\mu}
\end{equation}
the sum running over the set $\set S$ of inequivalent irreducible
representations ({\bf irreps}) of $\grp G$ contained in the decomposition of
$\sU_\grp G$, $\sH_\mu (\Cmplx^{m_\mu})$ being the representation
(multiplicity) space of the irrep $\mu$, of dimension $\dim (\sH_\mu) = d_\mu$
($\dim \Cmplx^{m_\mu} = m_\mu$).  The {\bf group average} of an
operator $O \in \L {\sH_\inp}$ is then given by
\begin{equation}\label{AveOp}
  \<O\>_{\sU_\grp{G}}:= 
  \int_\grp G \d g ~ U_g O U_g^\dag = \bigoplus_{\mu \in \set S} d^{-1}_{\mu} \left(\openone_{d_{\mu}} \otimes  \Tr_{\sH_\mu}[ \Pi_{\mu}O \Pi_\mu] \right),
\end{equation}
$\Pi_\mu$ denoting the projector onto $\sH_\mu\otimes\Cmplx^{m_\mu}$.
For the dilation we introduce now two  ancillary spaces
$\sH_0 \simeq\Cmplx^r$  where $r=\rnk {(S_0)}$, and $\tilde{\sH}$, which is given by
\begin{equation}
  \tilde {\sH} := \bigoplus_{\mu \in \set S}\sH_\mu\otimes\Cmplx^{d_\mu}~,
\end{equation}
and carries the representation $\tilde\sU_\grp{G}:=\{\tilde{U}_g=
\bigoplus_{\mu \in \set S} U_g^\mu\otimes \openone_{d_\mu} \} $.
\begin{Prop}\label{l:pi} 
  Let $\map{Z}$ be an instrument with outcome space $\Omega =\grp
  G/\grp G_0$, covariant w.r.t.
  $\{\grp{G},\mU_{\grp{G}},\mV_{\grp{G}}\}$, and with density $\map
  S_\omega$. A dilation of $\map Z$ can be achieved as follows
  \begin{equation}\label{Spi}
    \map{S}_{\pi(g)} (\rho) = V_g~\left( \Tr^{\phantom {H^H}}_{\sH_0}\Tr_{\tilde\sH}\left[(\openone_{\sH_0} \otimes \openone_\out
      \otimes \zeta'_g)~V' \rho V'^\dag\right] \right)~V_g^\dag ,
  \end{equation}
  where $\pi:g\mapsto\pi(g)\in\grp{G}/\grp{G}_0$ projects group
  elements to the corresponding left coset,
 $\zeta'_g $ is the POVM density on $\widetilde \sH$ given by 
\begin{equation}\label{covzeta}
\zeta'_g=|\eta_g\>\<\eta_g| \qquad   |\eta_g\>= \bigoplus_{\mu\in\set S}\sqrt{d_{\mu}}|U_g^\mu \kk \in \widetilde \sH,
  \end{equation}
and  $V': \sH_\inp \to \sH_0\otimes\sH_\out\otimes\tilde{\sH}$ is the isometry
  \begin{equation}
    V'  =\sum_{i=1}^r|i\> \otimes\int_\grp{G} \d g~ S_i U_g^\dag \otimes |\eta_g\>,
  \end{equation} 
 $\{|i\>\}_{i=1}^r$ being an orthonormal basis for $\sH_0$.
\end{Prop}
\Proof 
As an immediate consequence of Eq.
(\ref{AveOp}), the vectors $|\eta_g\> = \tilde U_g |\eta\>$
provide a resolution of the identity in $\widetilde\sH$, namely
\begin{equation}
\int_\grp{G} \d g~ |\eta_g\>\<\eta_g| =\openone_{\widetilde\sH}~,
\end{equation}
whence $\zeta'_g$ is the density of a normalized POVM.  Moreover, it is easy verify that $V'$ is an isometry. First, we have $\<\eta_g|\eta_h\> = \sum_{\mu
  \in \set S}~ d_{\mu}~ \chi_{\mu}(g^{-1}h)$, where
$\chi_{\mu}(g)\doteq \Tr[U_g^{\mu}]$ is the character of the irrep
$\mu$. Then, as a consequence of the orthogonality of irreducible
matrix elements, we have the relation
\begin{equation}\label{IntRep}
  \int_\grp{G} \d g~ \left( \sum_\mu d_{\mu} \chi_{\mu}^* (g)\right)~ U_g = \openone_\inp~,
\end{equation}
whence
\begin{equation}
  \begin{split}
    V'^\dag V' =& \sum_i \int_\grp{G} \d g \int_\grp{G} \d h~ U_g S_i^\dag S_i U_h^\dag ~ \<\eta_g|\eta_h\>\\
    =& \int_\grp{G} \d g \int_\grp{G} \d h~ U_g \xi U_h^\dag \left(\sum_\mu d_{\mu}~ \chi_{\mu}(g^{-1}h) \right)\\
    =& \int_\grp{G} \d g \int_\grp{G} \d k~ U_g \xi U_k U_g^\dag
    \left( \sum_\mu~ d_{\mu} \chi^*_{\mu}(k) \right)=\int_\grp{G} \d
    g~ U_g \xi U_g^\dag = \openone_\inp,
  \end{split}
\end{equation}
having used Eq. (\ref{frameorbitnorm}) with $\sA_\Omega \equiv \sU_{\grp G}$.
Finally,  identity (\ref{Spi}) holds, namely
\begin{equation}
  \begin{split}& V_g \left( \Tr_{\sH_0}^{\phantom{H^H}}\Tr_{\tilde\sH} \left[V' \rho V'^\dag (\openone_A\otimes\openone_\sH\otimes
        \zeta'_g)\right] \right) V^\dag_g =\\ =&   V_g \left( \sum_{i=1}^r \int_\grp{G} \d h \int_{\grp
        G} \d k~ S_i U_h^\dag \rho U_k S_i^\dag ~ \<\eta_g|\eta_h\> ~
      \<\eta_k|\eta_g\> \right) V_g^\dag\\
    =& V_g \left( \sum_{i=1}^r \int_\grp{G} \d h \int_\grp{G} \d k~ S_i U_h U_g^\dag 
      \rho U_g U_k S_i^\dag \left( \sum_{\mu \in \set S} d_{\mu}
        \chi_{\mu}^*(h)\right) ~ \left( \sum_{\nu \in \set S} d_{\nu} \chi^*_{\nu}
        (k)\right)\right) V_g^\dag\\ =& V_g \left(\sum_{i=1}^r S_i U_g^\dag \rho U_g S_i^\dag \right) V_g^\dag = \map V_g \map S_0 \map U_g^\dag (\rho)= \map S_{\pi(g)} (\rho)~,
  \end{split}
\end{equation}
having used Eqs. (\ref{IntRep}) and (\ref{covdens}).
\qed

The above proposition shows that in order to realize the instrument
$\map Z$ it is enough to perform the indirect measurement
$\zeta_g$---whose outcome space is the whole group $\grp G$---and
subsequently to use the classical data-processing
$g\longmapsto\pi(g)$, that projects the $g$ onto the final outcome
space $\Omega=\grp G/\grp G_0$. In this way, both the statistics and
the state reduction associated to the operational scheme of
measurement and feed-forward are exactly the same as for the
instrument $\map{Z}$.

\subsubsection{Naimark dilation} Consider the Hilbert space
$\bigoplus_{\mu \in \widehat{\grp G}} \sH_{\mu} \otimes
\Cmplx^{d_{\mu}}$ where $\widehat{\grp G}$ denotes the set of all possible unitary
irreps of $\grp G$. According to Fourier-Plancherel theory
\cite{Folland1995}, any vector $|f\> \in \bigoplus_{\mu \in \widehat{\grp G}} \sH_\mu \otimes
\Cmplx^{d_\mu}$ is associated with a square-summable function $f(g)$ as follows
\begin{equation}
|f\> \mapsto  f(g) = \sum_{\mu \in \widehat{\grp G}}  \sqrt{d_\mu}  \bb U_g| \Pi_{\mu} |f\>~.
\end{equation}
In this way, one has $\<f|h\> = \int_{\grp G} \d g ~ f^*(g) h(g)$, and
correspondence $|f\> \mapsto f(g)$ sets up a unitary equivalence
between the Hilbert spaces $\bigoplus_{\mu \in \widehat{\grp G}}
\sH_\mu \otimes \Cmplx^{d_\mu}$ and $L^2 (\grp G,\d g)$.  Therefore,
we can identify the ancilla space $\widetilde{\sH}$ in proposition
\ref{l:pi} with a subspace of $L^2 (\grp G, \d g)$, the projector on
$\widetilde{\sH}$ being $Y =\bigoplus_{\mu \in \set S} \Pi_\mu$.
Hence the POVM $Q'$ defined by the density $\zeta'_g$ in Eq.
(\ref{covzeta}) has the following Naimark dilation
\begin{equation}
\begin{split}
  \<f| Q'_B|h\> &= \int_B \d g ~\sum_{\mu,\nu \in \set S} ~ \sqrt{d_\mu
    d_\nu} ~ \<f|\Pi_\mu | U_g^\mu \kk \bb U_g^\nu| \Pi_\nu |h\>\\
&= \int_B \d g ~  (Y f)^* (g)  (Y  h)(g)\\
&= \<f | Y^\dag E_B Y |h\> \qquad \forall |f\>, |h\> \in L^2 (\grp G, \d g)~,  
\end{split}
\end{equation}
where $E$ is the PVM on $L^2 (\grp G, \d g)$  defined by 
\begin{equation}
\<f|E_B |h\> := \int_B \d g ~ f^*(g) h(g) \qquad \forall |f\>,|h\> \in L^2 (\grp G, \d g)~.
\end{equation}
The relation $Q_B = Y^\dag E_B Y$ shows that the POVM $Q'$ is simply
the projection of the PVM $E$ on the subspace $\widetilde{\sH} \subset
L^2 (\grp G, \d g)$. It is worth noting that the POVM $Q'$ is also the
optimal POVM for the estimation of an unknown unitary transformation
$\tilde U_g$ acting on the finite dimensional Hilbert space
$\widetilde \sH$ \cite{Chiribella:2005p3167}.

\section{Tight operator frames: Tele-instruments and
  Tele-channels}\label{tele}
Let $\mu$ be a finite measure on $\Omega$ and $\sA_\Omega$ a
measurable family of operators. $\sA_\Omega$ is a tight operator frame
if the frame operator $A$ is the identity on $\sH^{\otimes 2}$, i.e.
\begin{equation}
  A =\int_\Omega \mu(\d \omega) |A_\omega \kk \bb A_\omega | = \openone \otimes \openone
\end{equation}
A special case of tight unitary frame is that of irreducible unitary
representation of a compact group $\grp{G}$, namely $\Omega = \grp G
\, , \sA_\Omega = \sU_{\grp G}$ and $\mu (\d \omega)= \dim {\sH}
\times \d g$.

Generalizing the notion of tight frame to the case where the frame
operator is the identity only on the first copy we have the following
\begin{Def} 
  Let $\mu$ be a measure on $\Omega$ and $\sA_\Omega$ be a measurable
  family of operators. We say that $\sA_\Omega$ is a {\bf left-tight
    operator frame} if
  \begin{equation}\label{tframe}
    \int_\Omega \mu(\d\omega) \,  |A_\omega \kk \bb A_\omega |=  \openone \otimes  K
  \end{equation}
for some positive operator $0\le K \in \L{\sH}$.
\end{Def}

Note that identity (\ref{tframe}) is equivalent to the following ones
\begin{equation}\label{tframe2}
  \begin{split}
    \int_\Omega \mu(\d\omega) \, A_\omega X A_\omega^\dag&= \Tr[X K^\tau] 
    ~\openone,
    \quad\forall X\in\L{\sH} \\
    \int_\Omega \mu(\d \omega) \, A_\omega \otimes A_\omega^* &=
    |\openone \kk \bb K^\tau|~.
  \end{split}
\end{equation}

\bigskip 

We will see in the following that operator-frame instruments generated
by operations $\oA_\omega(\cdot)=A_\omega\cdot A_\omega^\dag$
corresponding to left-tight frames $\sA_\Omega$ can be realized by
{\bf generalized teleportation schemes}, in which two parties (conventionally
named Alice and Bob) exploit an entangled resource to achieve the
instrument via local operations and one-way classical communication: a
suitable joint POVM $\zeta_\omega$ is measured by Alice on the input
and on one side of the entangled state, the measurement outcome
$\omega$ is announced to Bob, who performs a conditional feed-forward
operation $\map B_\omega$ on the other side.  We will also use the
term {\bf tele-instruments} to denote instruments that admit such a
realization. In addition, we will show that frame-orbit instruments
generated by tight unitary frames are useful for the realization of
covariant channels. In this case, which covers in particular the case
of unitary irreducible group representations, a covariant channel can
be realized by a generalized teleportation scheme, hence becoming a {\bf
  tele-channel}.  In particular, we will provide also the realization
of covariant channels such as universal cloning \cite{werner98} and
universal NOT \cite{opttransp}.

\subsection{Minimal tele-instruments}\label{s:mintel}
Let $\map Z : \Sigma_\Omega \to \CP {\sH_\inp, \sH_\out}$ be a
frame-orbit instrument w.r.t. $(\mu, \oA_\Omega, \sB_\Omega)$
with $\oA_\Omega$ left-tight operator frame, and let $\map S_\omega$
be the instrument density of $\map Z$.  We consider now the minimal
feed-forward realization of Corollary \ref{Cor:feed-for}.  Using Eq.
(\ref{tframe2}), the CJ operator of the instrument channel
$\map{T}_\Omega$ is given by
\begin{equation}
\begin{split}
  T_\Omega=&(\map{T}_\Omega\otimes\map{I})(|\openone_\inp\kk\bb\openone_\inp
  |)= \int_\Omega \mu(\d\omega)~( \openone_\out \otimes  A^*_\omega 
   )S_0 (\openone_\out \otimes A^\tau_\omega)\\=&
   \Tr_\inp [ S_0 (\openone_\out \otimes (K^*)^\tau )] \otimes \openone_\inp \\  = &
   \Tr_\inp [ S_0 (\openone_\out \otimes K)] \otimes \openone_\inp \\
=&  \sigma \otimes \openone_\inp  ~,
\end{split}
\end{equation}
having used the fact that $(K^*)^\tau = K$ since $K\ge 0$, and having defined the state 
\begin{equation}\label{sigmatight}
\sigma := \Tr_\inp [S_0 (\openone_\out \otimes K)]= \map S_0 (K^\tau) \qquad \Tr [\sigma] =1~.
\end{equation}

The minimal isometry $V = (\openone_\out \otimes {T_\Omega^{\tau \frac12}}) |\openone_\out\kk \otimes \openone_\inp$ is then given by:   
\begin{equation}
V  =  |{\sigma}^\frac12 \kk \otimes \openone_\inp~ 
\end{equation}
and Corollary \ref{Cor:feed-for}  yields the following
generalized teleportation scheme for the instrument $\map Z$:
\begin{equation}\label{minimaltight}
  \map S_\omega (\rho)= \map B_\omega \left( \Tr_{2,3}[(|{\sigma}^\frac12\kk\bb{\sigma}^\frac12|\otimes\rho)(\openone_\out\otimes
  \zeta_\omega)] \right)  
\end{equation}
where the POVM density $\zeta_\omega$ is given by
  $\zeta_\omega = (\sigma^{-\frac12 \ \tau} \otimes A_\omega) S_0^\tau (\sigma^{-\frac12 \ \tau} \otimes A_\omega^\dag)$.

  In conclusion, in the minimal feed-forward realization the
  frame-orbit instrument $\map{Z}$ can be implemented by two parties
  that share the pure entangled state $|{\sigma}^\frac12 \kk$ by using
  only local operations and one-way classical communication: it is
  enough for Alice to perform the joint POVM $\zeta_{\omega} =
  (\sigma^{-\frac12 \ \tau} \otimes A_\omega) S_0^\tau (\sigma^{-\frac12 \ \tau}
  \otimes A_\omega^\dag) $ on the input state and on one side of the
  entangled resource, and to announce the measurement outcome $\omega$
  to Bob, who implements the conditional channel $\map B_\omega$.

\subsection{Non-minimal tele-instruments}\label{s:maxtel}
Starting from the minimal dilation it is simple to obtain other generalized teleportation realization schemes. In particular, from Eq. (\ref{minimaltight}) we obtain
\begin{equation}\label{nonminimaltight}
\begin{split}
  \map S_\omega &= \map B_\omega \left ( \Tr_{2,3} \left[
      \left(|\sigma^\frac12\kk\bb\sigma^\frac12|\otimes\rho \right)
      \left(\openone_\out\otimes (\sigma^{-\frac12 \ \tau} \otimes
        A_\omega) S_0^\tau (\sigma^{-\frac12 \ \tau}
        \otimes A_\omega^\dag ) \right) \right]\right) \\
  &= \map B_\omega \left( \Tr_{2,3} \left[ \left(|\openone_\out\kk\bb
        \openone_\out|\otimes\rho \right) \left(\openone_\out\otimes
        (\map S_0^\tau \otimes \map I) ( |A^\tau_\omega \kk \bb  A^\tau_\omega| ) \right) \right] \right) \\
  &= \map B_\omega \left( \Tr_{2,3} \left[
      \left( (\map I \otimes \map S^\tau_0)(|\openone_\out\kk\bb \openone_\out|) \otimes\rho \right) \left(\openone_\out\otimes |A^\tau_\omega \kk \bb A^\tau_\omega |  \right) \right] \right) \\
  &= \map B_\omega \left( \Tr_{2,3} \left[
      \left( (\map S_0 \otimes \map I)(|\openone_\out\kk\bb \openone_\out|) \otimes\rho \right) \left(\openone_\out\otimes |A^\tau_\omega \kk \bb A^\tau_\omega|  \right) \right] \right) \\
  &= \map B_\omega \left( \Tr_{2,3} \left[
      \left( (\map S_0 \otimes \map I)(|K^{\tau \frac12}\kk\bb K^{\tau \frac12}|) \otimes\rho \right) \left(\openone_\out\otimes |K^{-\frac12} A^\tau_\omega \kk \bb K^{-\frac12} A^\tau_\omega|  \right) \right] \right) \\
  &=\map B_\omega \left( \Tr_{2,3} \left[ \left( \sigma' \otimes\rho
      \right) \left(\openone_\out\otimes
        \zeta'_\omega  \right) \right] \right) \\
\end{split}
\end{equation} 
having defined the state 
\begin{equation}
  \sigma' := (\map S_0 \otimes \map I)(|K^{\tau \frac12}\kk\bb K^{\tau \frac12}|)
\end{equation}
and the POVM density 
\begin{equation}
\zeta_\omega' := | K^{-\frac12} A^\tau_\omega \kk \bb
K^{-\frac12} A^\tau_\omega |~.
\end{equation} 
The normalization of $\sigma'$ is given by $\Tr[\sigma'] = \Tr [\map
S_0 (K^\tau)] = \Tr[\sigma]=1$ (having used Eq. (\ref{sigmatight})),
while the normalization of the POVM $\zeta_\omega'$ follows directly
from Eq. (\ref{tframe}).

Also the above feed-forward realization is a generalized teleportation
scheme, which allows Alice and Bob to implement the instrument $\map
Z$ as a tele-instrument. Notice that now the entangled resource
$\sigma'= (\map S_0 \otimes \map I)(|K^{\tau \frac12}\kk\bb K^{\tau \frac12}|) $
is a generally \emph{mixed} state, whereas the joint POVM performed by
Alice has now the \emph{rank-one} density $\zeta_{\omega}' = |K^{-\frac12}
A^\tau_\omega \kk\bb K^{-\frac12} \transp A_\omega |$.

\subsection{Tight unitary frames and Bell measurements}
A particularly interesting generalized teleportation scheme arises when the
left-tight frame $\sA_\Omega$ consists of unitary operators, namely
$\sA_\Omega \equiv \sU_\Omega$.  This is the case, for instance, when
$\sA_\omega$ is a unitary irreducible representation of some compact
group $\grp G$.

It is immediate to see that a unitary left-tight frame $\sA_\Omega$ is
is necessarily tight, since Eq. (\ref{tframe}) implies $K=
\openone_\inp /d_\inp$. For unitary tight frames the non-minimal
realization of Eq.(\ref{nonminimaltight}) becomes
\begin{equation}\label{nonminimalunitary}
  \map S_\omega =\map B_\omega \left(  \Tr_{2,3} \left[
      \left( \frac{S_0 }{d_\inp} \otimes\rho \right) \left(\openone_\out\otimes
        \zeta'_\omega  \right) \right] \right)  \qquad \zeta'_\omega = d_\inp | \transp U_\omega \kk \bb \transp U_\omega|~, 
\end{equation}
and the joint POVM used by Alice is a Bell measurement, i.e.
$\zeta'_\omega$ are proportional to rank-one projectors on maximally
entangled states. On the other hand, the minimal realization of Eq.
(\ref{minimaltight}) gives
\begin{equation}\label{minimalunitary}
  \map S_\omega (\rho)= \map B_\omega \left( \Tr_{2,3}[(|\sigma^\frac12\kk\bb\sigma^\frac12|\otimes\rho)(\openone_\out\otimes
    \zeta_\omega)] \right)  
\end{equation}
with $\sigma = \map S_0 (\openone_\inp) /d_\inp$, and POVM density
$\zeta_\omega = (\sigma^{-\frac12 \ \tau} \otimes U_\omega) S_0^\tau
(\sigma^{-\frac12 \ \tau} \otimes U_\omega^\dag)$.  Note that typically the
POVM $\zeta_\omega$ in the minimal realization is not a Bell POVM.

\section{Realization of  covariant channels}\label{sec:covchan}
The realization of a covariant instrument $\map{Z}$ also allows one to
achieve the corresponding channel $\map{Z}_\Omega$ by simply averaging
over the instrument outcomes. Therefore, the general realization
schemes presented in Theorem \ref{t:dilations} and Corollary
\ref{Cor:feed-for} for instruments can be directly transferred to the
corresponding channels.  In particular, for any tele-instrument $\map
Z$ in subsections \ref{s:maxtel} and \ref{s:mintel} we have a
corresponding tele-channel $\map Z_\Omega$ achieved by the same
generalized teleportation scheme.

A particularly interesting case is that of {\bf covariant channels},
which we intend here in a very broad sense, according to the
following:
\begin{Def}
  Let $\oA_\Omega$ be a family of quantum channels on $\sH_\inp$ and
  $\sB_\Omega$ a family of quantum channels on $\sH_\out$.  A channel
  $\map C \in \CP {\sH_\inp, \sH_\out}$ is \emph{covariant} w.r.t.
  $(\oA_\Omega, \sB_\Omega)$ if
\begin{equation}
  \map C \map A_\omega = \map B_\omega \map C \qquad \forall \omega \in \Omega~.
\end{equation}
\end{Def}
In particular we consider the case where all channels in $\oA_\Omega$
are unitary, namely $\map A_\omega(\cdot) \equiv \map U_\omega (\cdot)
= U_\omega \cdot U_\omega^\dag$, for some unitary operator $U_\omega \in \L{\sH_\inp}$.  
Since for a covariant channel one has $\map C= \map B_\omega \map
C\map U_\omega^\dag \ \forall \omega\in \Omega$, any covariant channel
is trivially the channel corresponding to a frame-orbit instrument,
namely $\map C \equiv \map Z_\Omega$ with $\map Z_B := \int_B \mu (\d
\omega) \map S_\omega , \ \map S_\omega : = \map B_\omega \map C\map
U_\omega^\dag$. In fact, the covariant channel $\map C$ coincides with
the instrument density $\map S_\omega$ for any outcome $\omega$.  In
particular, when $\sU_\Omega$ is a tight unitary frame, $\map C$
becomes a tele-channel, and the non-minimal dilation of Eq.
(\ref{nonminimalunitary}) yields a generalized teleportation scheme
with Bell measurement:
\begin{equation}\label{nonminimalcovariant}
  \map C(\rho) \equiv \map S_\omega (\rho) =  \map B_\omega \left( \Tr_{2,3} \left[ \left(  C/{d_\inp} \otimes \rho \right)  \left( \openone_\out \otimes d_\inp |\transp U_\omega \kk \bb \transp U_\omega | \right)\right]\right)~, 
\end{equation}
with $C   =(\map C \otimes \map I) (|\openone_\inp\kk \bb
\openone_\inp |) \equiv  (\map S_0 \otimes \map I) (|\openone_\inp\kk \bb
\openone_\inp |) = S_0$.

The minimal dilation of Eq. (\ref{minimaltight}) gives instead 
\begin{equation}\label{minimalcovariant}
 \map C(\rho) \equiv \map S_\omega (\rho)= \map B_\omega \left( \Tr_{2,3}[(|\sigma^\frac12\kk\bb\sigma^\frac12|\otimes\rho)(\openone_\out\otimes
  \zeta_\omega)] \right)  
\end{equation}
where $\sigma = \map C(\openone_{\inp})/d_\inp$ according to Eq. (\ref{sigmatight}), and POVM density given by $\zeta_\omega = (\sigma^{-\frac12 \ \tau} \otimes U_\omega) C^\tau (\sigma^{-\frac12 \ \tau} \otimes U_\omega^\dag)$.

  Notice
that the non-minimal realization uses the Choi-Jamio\l kowski state
$\sigma_{CJ}= C/d_\inp$ as entangled resource, while the minimal realization uses a purification of the local state $\sigma = \Tr_{\inp} [\sigma_{CJ}]$. The two realizations coincide (up to local unitaries on $\sH_\inp$) when the CJ-operator $C$ is rank-one, corresponding to unitary channels.

\medskip
We conclude with the following examples of application:
\subsubsection{Ideal teleportation} Ideal teleportation from Alice's
to Bob's site is described by the identity channel $\map C= \map I$,
which is a covariant channel w.r.t. $(\mU_\Omega, \mU_\Omega)$
for \emph{any} unitary frame $\sU_\Omega$, since trivially $\map C
\map U_\omega = \map U_\omega \map C \ \forall \omega \in \Omega$. For
tight unitary frames, Eqs. (\ref{minimalcovariant}) and
(\ref{nonminimalcovariant}) coincide and give the realization
\begin{equation}
  \map C (\rho) =\map U_\omega \left(  \Tr_{2,3} \left[
      \left( \frac{|\openone \kk\bb \openone|}{d} \otimes\rho \right) \left(\openone_\out\otimes  d | \transp U_\omega \kk \bb \transp U_\omega|
      \right) \right] \right)  ~.
\end{equation}
 In other
words, our general scheme retrieves all possible schemes to achieve
ideal teleportation with Bell observables \cite{Bennett:1993p3168,Werner:2001p3169},
and, more generally, with Bell POVMs \cite{Braunstein:2000p3170}.

\subsubsection{Universal tele-cloning.} The optimal quantum cloning of
pure states from $N$ input copies to $M$ output copies is given by a
channel $\map C_{N,M} \in \CP {\sH_N^+, \sH_M^+}$, where $\sH_k^+$
denotes the totally symmetric subspace of the tensor product
$\sH^{\otimes k}$. The channel is covariant w.r.t. the irreducible
representations $(\mU_\grp G^{\otimes N}, \mU_{\grp G}^{\otimes M})$
of the group $\grp G = \mathbb {SU} (d)$, namely $\map C_{N,M} \map
U_g^{\otimes N} = \map U_g^{\otimes M} \map C_{N,M} \ \forall g \in
\mathbb {SU} (d)$, and is given by \cite{werner98,Keyl:1999p3171}
\begin{equation}
\map C_{N,M} (\rho)  = \frac{d_N^+} {d_M^+} ~  P_M^+ (\rho \otimes \openone^{\otimes (M-N)}) P_M^+ \qquad \rho \in \St {\sH_N^+},
\end{equation} 
where $d_k^+ := \dim {\sH_k^+}$, and $P_k^+$ is the projector on
$\sH_k^+$. The realization of Eq.  (\ref{nonminimalcovariant}) then
yields a generalized teleportation scheme with covariant Bell POVM:
\begin{equation}
  \map C_{N,M} (\rho) =\map U_g^{\otimes M} \left(  \Tr_{2,3} \left[
      \left( \frac{C_{N,M} }{d_N^+} \otimes\rho \right) \left(\openone_\out\otimes
        d_N^+| U_g^{\tau \otimes N } \kk \bb U_g^{\tau \otimes N}|  \right) \right] \right)   , 
\end{equation}
with $C_{N,M} = (\map C_{N,M} \otimes \map I) (|\openone_{\sH_N^+} \kk \bb
\openone_{\sH_N^+}|) $.
On the other hand, the feed-forward scheme of Eq. (\ref{minimalcovariant}) gives
\begin{equation}
\map C_{N,M} (\rho) = \map U^{\otimes M}_g \left (\Tr_{2,3} \left[ \left( \frac{|P_M^+ \kk \bb P_M^+|}{d_M^+} \otimes \rho \right) \left( \openone_\out \otimes  \zeta_g \right)\right]  \right)
\end{equation}
where $\zeta_g$ is the covariant POVM given by
\begin{equation}
\zeta_g ={d^+_M} {(\openone_\out \otimes U_g^{\otimes N})  C^\tau_{N,M}(\openone_\out \otimes U_g^{\dag \otimes N})} 
\end{equation}

\subsubsection{Optimal universal NOT gate} The optimal universal NOT
is the channel from $\sH_N^+$ to $\sH_M^+$ with $\sH = \Span
\{|0\>,|1\>\} \simeq \Cmplx^2$ which transforms $N$ copies of a pure state into one approximate copy of its orthogonal complement. The channel $\map N$ is given by the measure-and-reprepare scheme \cite{Ricci:2004p3164}
\begin{equation}
 \map N (\rho) = \int_{\mathbb {SU} (d)}\!\!\!\!\!\! \d g ~  \Tr [\rho  \zeta_g]~ U_g |1\>\<1| U_g^\dag~. 
\end{equation}
where $\zeta_g$ is the covariant POVM $\zeta_g = d_N^+ ~\left(U_g
  |0\>\<0|U_g^\dag\right)^{\otimes N}$. By definition, $\map N
\equiv \map Z_\Omega$, where $\map Z$ the covariant channel with
density $\map S_g (\rho) = \Tr [\rho \zeta_g] ~ U_g|1\>\<1|U_g^\dag =
\map U_g \map S_0 \map U_g^{\dag \otimes N} , \ \map S_0 (\rho) =
\Tr[\rho (|0\>\<0|)^{\otimes N}]~ |1\>\<1|$. In this case, it easy to see that the minimal
generalized teleportation scheme given by Eq. (\ref{minimaltight}) coincides
with the definition of the channel: indeed $\map N$ is of
the measure-and-reprepare form, and by definition it can be achieved via a measurement at Alice's site combined with a conditional state preparation at Bob's site. 
On the other hand, the non-minimal scheme of Eq. (\ref{nonminimaltight}) gives 
\begin{equation}
\map N (\rho) = \int_{\mathbb {SU}(d)}\!\!\!\!\!\! \d g ~\map U_g \left(\Tr_{2,3}[\left (|1\>\<1| \otimes |0\>\<0|^{\otimes N}  \otimes \rho  \right) \left( \openone_\out \otimes   d_N^+ |U_g^{\tau \otimes N } \kk \bb U_g^{\tau \otimes N }|   \right) ] \right).
\end{equation}

\bibliographystyle{unsrt}

\end{document}